# Lattice and Magnetic Effects on a *d-d* Excitation in NiO Using a 25 meV Resolution X-ray Spectrometer

Daisuke Ishikawa[1,2], Maurits W. Haverkort[3] and Alfred Q. R. Baron[1]

[1]*Materials Dynamics Laboratory, RIKEN SPring-8 Center, 1-1-1 Kouto, Sayo, Hyogo, 6791-5148 Japan*
[2]*Research and Utilization Division, SPring-8/JASRI, 1-1-1 Kouto, Sayo, Hyogo 679-5198 Japan*
[3]*Institute for Theoretical Physics, University of Heidelberg, Seminarstrasse 2, 69117 Heidelberg, Germany*

We investigate the behavior of a *d-d* transition in NiO using a new x-ray spectrometer with 0.025 eV resolution at 15816 eV, and via *ab-initio* ligand field theory calculations. The transition at ~1.7 eV energy transfer is measured at temperatures between 20 and 800 K, at a momentum transfer |**Q**| = 6.52 Å$^{-1}$. Fine structure is clearly observed at 20 K. As temperature is increased, the excitation shifts to lower energy and broadens. We explain the energy shift as being related to thermal expansion and to magnetism. The broadening is well fit considering thermal fluctuations of the Ni-O bond length, with a scale factor found to be in reasonable agreement with calculation.

The electronic spectroscopy of materials is of fundamental interest for understanding their properties. In principle, non-resonant inelastic x-ray scattering (NRIXS) is an excellent tool for such spectroscopic investigations, with access to large momentum transfers allowing investigation beyond the dipole limit, and over atomic scale correlation lengths. However, there are significant practical constraints. Previously, NRIXS has been a powerful tool for studying electronic[1] and atomic[2] dynamics. While atomic dynamics can be studied with extremely high, ~meV, resolution, as the cross section is relatively large, the cross section for electronic excitations is much smaller, so those measurements require relaxation of energy and/or momentum resolution to obtain sufficient data rate.[3] NRIXS with ~eV resolution has been used to investigate valence structure, particle-hole excitations and plasmons,[1] however it is only relatively recently that experimental work has begun to focus on narrower electronic (non-phonon) spectral features at ~eV energy transfers.[4] This, in large part, is due to instrumental limitations, as one needs both high resolution and high spectral intensity. However, given the possibility to directly investigate transitions of d electrons and the dispersion of such excitations (e.g., orbitons) there is significant interest.

Non-resonant scattering offers some advantages relative to resonant inelastic x-ray scattering (RIXS) (see, e.g., Refs 1, 5 and 6). NRIXS is *relatively* simple to interpret and calculate as the NRIXS cross section responds directly to the electronic charge density of the initial and final states via the **A**$^2$ term of the interaction Hamiltonian: avoiding the intermediate states from the resonant interaction, the **A** · **p** term, reduces count-rates but simplifies interpretation. The choice of x-ray energy in a non-resonant experiment is flexible, so it can be chosen to optimize optical performance to facilitate improved resolution. The energy can also be chosen to be relatively high to improve transmission into and out-of sample environments, and to increase the illuminated volume of a sample. Also, as compared to soft x-ray RIXS (also called SIXS) at ~1 keV, the high energy of NRIXS allows access to more momentum space.

Previous spectrometers for NRIXS have employed "dispersion compensation"[7] to improve the energy resolution, with recent work demonstrating operation at 50 to 60 meV resolution is possible for electronic excitations.[8,9] However, that approach requires placement of a position-sensitive detector very close to the sample, limiting the space for sample environments. Here we use a temperature gradient[10] to allow improved, 25 meV, resolution with a single element detector with a large space near the sample (Fig. 1(b)). We use this to investigate a band of *d-d* excitations in NiO as we increase the temperature from 20 to 800 K, through the Neel point at $T_N$ = 523 K.

The main elements of the spectrometer are shown in Fig. 1(a), and include the medium resolution monochromator, a bent cylindrical focusing mirror, stages and environment for the sample, and the analyzer system on the two-theta arm. This is installed at RIKEN BL43LXU[11] of SPring-8. The x-ray source is three tandem, 5 m long, in-vacuum insertion devices. A liquid-nitrogen-cooled vertically scattering flat mirror is used to reduce the beam power by removing harmonics. The bandwidth is reduced to ~2.8 eV using a liquid-nitrogen-cooled Si(111) high-heat load monochromator, and then a second mirror is used to return the beam to the horizontal. The operating energy of the spectrometer is 15816 eV, corresponding to the energy of the Si (888) back-reflection analyzer crystals.

The medium-resolution monochromator (MRM) is based on a "nested" design[12] as this is a relatively straightforward way of efficiently obtaining ~20 meV resolution at 15.8 keV. The outer, collimating, channel-cut crystal is a silicon (440) reflection with asymmetry factor *b* = -1/13. The inner crystal is a symmetric Si (660) reflection. Care was needed to compensate for the very high power load, ~1 W, of the 2.8 eV beam incident on the MRM: precision temperature control was applied to each blade of each channel-cut crystal. The resolution of the monochromator was measured to be 22 meV full width at half maximum (FWHM), in agreement with calculations, with a throughput of 2 THz. The elliptically bent cylindrical mirror after the MRM focused the beam to $20 \times 30$ μm$^2$, FWHM, at the sample location.

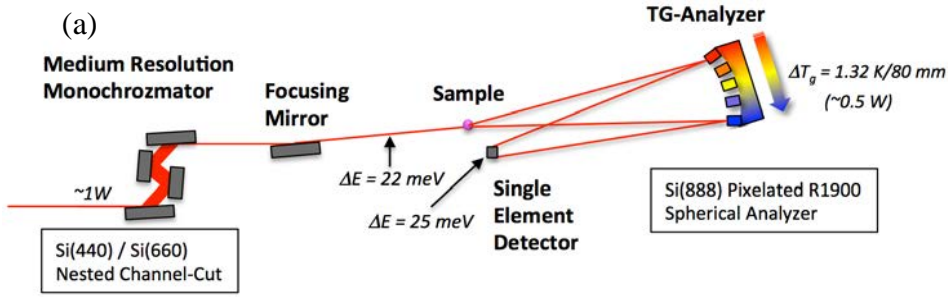 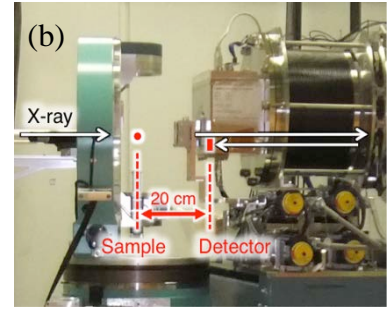

**Fig. 1.** 25 meV resolution TG spectrometer. (a) Schematic showing the main components. (b) Photo of the sample area showing the possible free space for the sample environment.

The analyzer crystal, fabricated using a 1.9 m radius spherically polished substrate, was diced to reduce strain (e.g., Ref. 13) following an extensive in-house R&D program.[14] A temperature-gradient (TG) of $\Delta T_g = 1.32$ K/80 mm was applied over the analyzer. The gradient compensates for variation in Bragg angles by changing the Si(888) d-spacing via thermal expansion, allowing the spectrometer resolution to be significantly improved (see inset in Fig. 2(a)), even using a single element detector placed 20 cm from the sample. This validates the TG concept of Ref. 10.

The NiO crystal, $1 \times 5 \times 10$ mm$^3$, was placed inside of a closed-cycle cryofurnace, which was inserted into the phi circle of a Huber 512.1 Eulerian cradle. The diameter of the window of the cryofurnace was 68 mm, comfortably within the ~200 mm diameter available around the sample. Measurements were done at **Q** = (2.5 2.5 2.5) reciprocal lattice units (rlu), |**Q**| = 6.52 Å$^{-1}$, with the acceptance of the analyzer being $\Delta$**Q** = (0.05 0.22 0.22) rlu. This **Q** is near the maximum intensity for this band of excitations.[4,15,16] Collection of the spectral window from 1200-2200 meV energy transfer required about 5 hours per temperature, with peak rates from the $d$-$d$ excitation of about 3 photons/s into the 0.0024 sr. solid angle of the analyzer. Spectra were measured at 8 temperatures between 20 and 800 K.

The measured spectra are shown in Fig. 3, after background subtraction. At low temperatures, some structure is evident, and the spectra can be fit using two Gaussian lines, or a more complex multiplet to reproduce all of the fine structure (however the $\chi^2$/degree of freedom become much less than unity, so the multiplet is speculative). At higher temperatures the response broadens, shifts to lower energies, and becomes increasingly Gaussian. The temperature dependence of the center[17] and width of the response are shown in Fig.

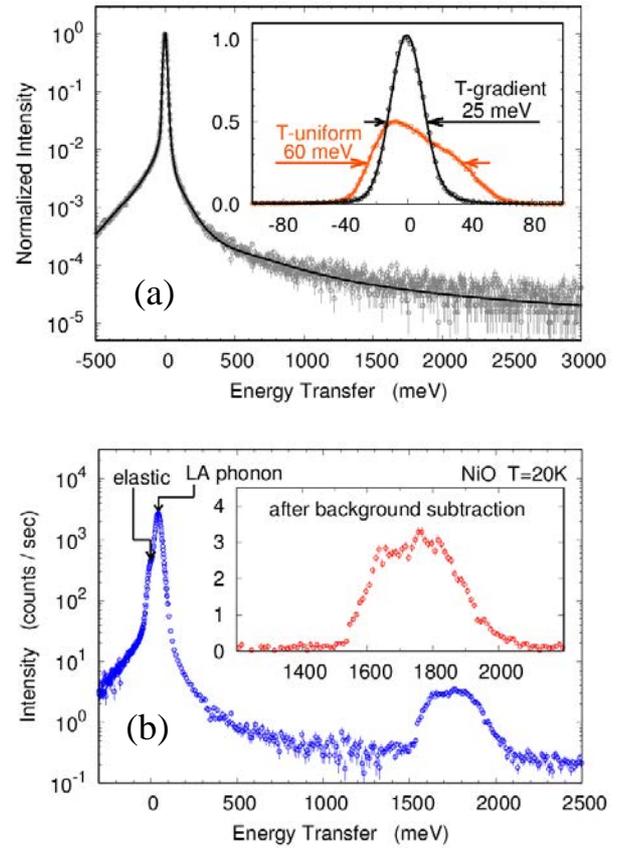

**Fig. 2.** Spectrometer response. (a) Measured scattering, with the TG, from plexiglass at its structure factor maximum where its response is predominantly elastic. The inset shows the resolution with and without TG. (b) The $d$-$d$ excitation in NiO appears clearly above background.

4. The energy decreases steeply in the neighborhood of the Neel point, and then continues to drop (Fig. 4(a)). The width increases approximately linearly with temperature (Fig. 4(b)).

Fine structure of the $d$-$d$ excitation in NiO is expected due to splitting of the transition from spin-orbit coupling and the non-spherical environment of the Ni ions. This environment is octahedral to a first approximation, but, below $T_N$, is further reduced in symmetry by the directionality of the AF ordering (the ordering is periodic along the (111) direction, with moments of neighboring Ni planes nearly antiparallel, and perpendicular to the ordering direction) and a concomitant weak rhombohedral distortion.[18-20] In the present case, one should average over different domains for the directionality of the AF order. Momentum-transfer resolved cluster calculations (see Refs. 15, 22 and discussion below), including the octohedral and rhombohedral distortion, tend to show two or three dominant lines, and a similar number of weak lines, with details that depend on both the averaging over domain structure, and the size of the assumed exchange interaction. The observed line-width (>300 meV) is larger than the line splitting of the cluster calculations (~150-200 meV), and similar to the width calculated for optical absorption spectra of this *d-d* band via Molecular Dynamics (MD) simulations.[21]

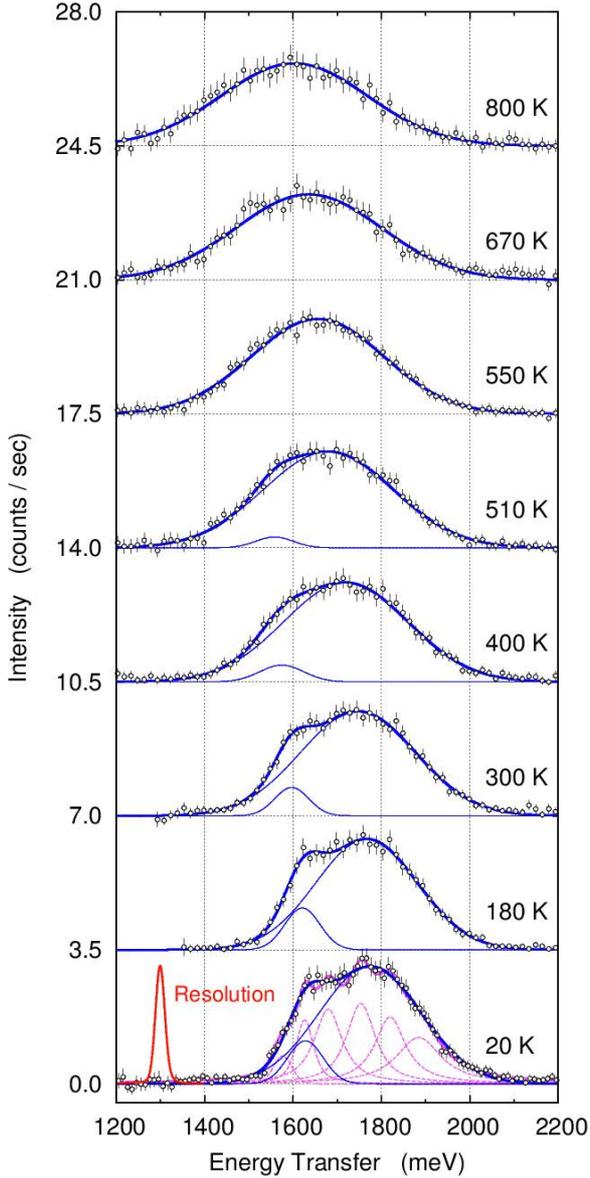

**Fig. 3.** Temperature dependence of the *d-d* spectra of NiO measured at Q = (2.5 2.5 2.5) rlu. One TG analyzer (0.0024 sr) and single element detector 2 × 2 mrm$^2$ were used. Data are shown after background subtraction. The resolution is shown in the lower left. The spectra are fit with two Gaussians below $T_N$ and a single Gaussian above $T_N$. The dashed lines show a possible fit with 6 lines, as gives nearly all the structure at the lowest temperature (see text).

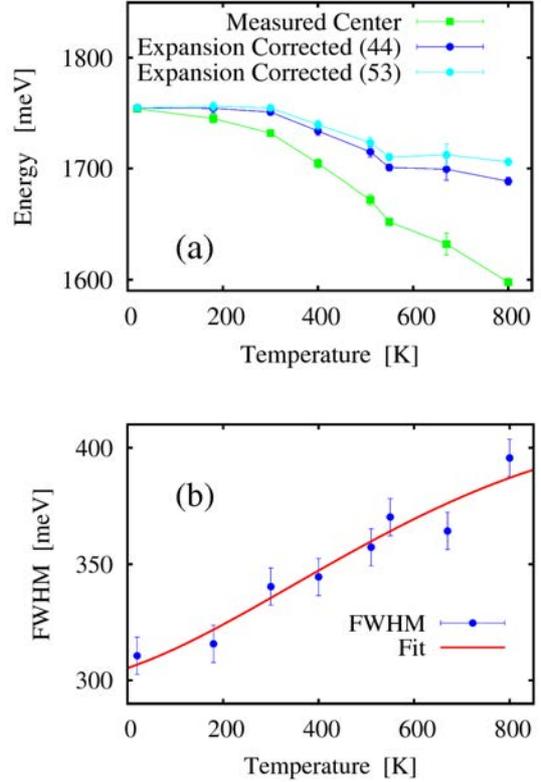

**Fig. 4.** Temperature dependence of the *d-d* excitation parameters. (a) Line position before and after correction for thermal expansion, using either the calculated value for the energy change (44 meV/pm) or the experimental value (53 meV/pm, see text). Solid lines are guides for the eye. (b) Width of the excitation compared to the model in the text.

Lattice effects, including both thermal expansion and vibrations, should be considered to understand our result: lattice expansion is expected cause to a decrease in excitation energy, while vibrations should give a temperature-dependent contribution to the line-width. The theoretical dependence of the crystal-field excitation energy on the NiO lattice parameter was calculated using *ab-initio* ligand field theory.[22] As a starting point we performed density functional theory (DFT) calculations, using both scalar and fully relativistic (with spin-orbit coupling) functionals.[23] The observed *d-d* transition energies are not equal to the onsite energy difference between the center of the $t_{2g}$ and $e_g$ bands (around ~1 eV) as the Coulomb energy between the Ni *d* orbitals of the initial and final state is different. The three possible *d-d* excitations are captured in a full many-body description that allows for entanglement between electrons described by multi Slater-determinant states. We downfolded the

DFT calculations to the Ni-d and O-p orbitals spanning the states from -10 to 2 eV and used this basis to create a ligand-field (NiO$_6$ cluster) model which is solved by exact diagonalization. This allows prediction of the energy and intensity of the experimental IXS spectra, including fine structure, with an absolute accuracy of about 10% of the *d-d* transition energy.[22] These calculations give an energy decrease of 1.27% per pm of lattice constant expansion (44.5 meV/pm of Ni-O bond-length increase at 1750 meV) for the *d-d* transitions investigated here, both with and without spin-orbit coupling.

Correcting the measured energy for the thermal expansion of NiO[24,25] gives the upper curves in Fig. 4(a). The excitation energy drops steadily between room temperature and the Neel point, before flattening out, suggesting the (thermal expansion-corrected) excitation energy is responding to the magnetism. Indeed, while complicated by domain structure issues, the cluster calculations suggest changes in the x-ray allowed transitions at this momentum transfer leads to an effective drop in average excitation energy of ~40 meV when the exchange coupling is reduced to zero, in reasonable agreement with the observation (~50-60 meV). One notes that the agreement is improved, and the high-temperature portion is flatter, if we use the experimental value (see below) of 53 meV/pm.

The impact of vibrations on the linewidth can be estimated using a harmonic phonon model for NiO[26, 27] by considering the fluctuation of the Ni-O bond-length. Straight-forward application of the methodology of Ref. 28 gives the mean-square fluctuation of the bond length to be

$$\left\langle \left[ \hat{\mathbf{x}} \cdot (\mathbf{u}_1 - \mathbf{u}_2) \right]^2 \right\rangle = \frac{\hbar}{2N_{\mathbf{q}}} \sum_{\substack{\mathbf{q} \\ \text{1st Zone}}} \sum_{\substack{j \\ \text{Modes}}} \frac{1}{\omega_{\mathbf{q}j}} \times \left| \hat{\mathbf{x}} \cdot \left( \frac{\mathbf{e}_{\mathbf{q}j1}}{\sqrt{M_1}} - \frac{\mathbf{e}_{\mathbf{q}j2}}{\sqrt{M_2}} \right) \right|^2 \coth\left( \frac{\hbar \omega_{\mathbf{q}j}}{2k_B T} \right) \quad (1)$$

where 1 and 2 refer to adjacent Ni and O atoms, and **u** their displacement from equilibrium positions, with the Ni-O nearest-neighbor bond directed along the $\hat{\mathbf{x}}$ -direction. Phonon frequencies and eigenpolarizations are given by $\omega_{\mathbf{q}j}$ and $\mathbf{e}_{\mathbf{q}j1}, \mathbf{e}_{\mathbf{q}j2}$. The sum is over $N_{\mathbf{q}}$ momentum transfers, **q**, in the first Brillouin zone and modes $j = 1...6$ at a temperature, $T$. While the form of Eq. (1) is similar to the thermal displacement of an atom from its average location (replace the squared term in the sum by $1/M$), it is quite different in practice. The difference between the mass-weighted polarization vectors of adjacent atoms means the Ni-O bond-length fluctuation will tend to be both be smaller than the thermal displacements, and more responsive to optical phonon modes. This is because the low-energy acoustic modes that dominate the thermal displacements from average positions have long wavelengths so have reduced impact on the distance between adjacent atoms.

We fit the line-width taking

$$W = 2.35 \sqrt{\sigma_0^2 + a^2 \left\langle \left[ \hat{\mathbf{x}} \cdot (\mathbf{u}_1 - \mathbf{u}_2) \right]^2 \right\rangle}$$

to relate the observed FWHM, $W$, to the calculated temperature dependent fluctuation, where $\sigma_0$ is a temperature-independent contribution due to line splitting and static disorder. One obtains a good fit (the red line in Fig. 4(b)) with $\sigma_0$ =127 (3) meV and $a$ =53 (3) meV/pm of NiO bond length. Given the simplicity of this model, we believe this value for $a$ is in reasonable agreement with the theory value of 44.5 meV/pm.

We compare the present results to the recent NRIXS study of the temperature dependence of *d-d* excitations in CuO from 10 to 320 K.[9] In that work, changes were measured in the band of excitations at 2 eV in CuO, with broadening of the FWHM similar to that observed here (increasing ~100 meV from 400 to 500 meV), but no shift in position through the Neel temperatures at ~230 K, and with a line-shape that had rather similar features at 10 and 320 K. However, the poorer resolution for the CuO work (×2.4 in energy and ×6 [29] in momentum) might have blurred out spectral changes. Huotari, *et al.*,[9] suggested that the linewidth change with temperature that they observe in CuO is due to changes in the band gap which is at an energy just below the *d-d* excitations. For NiO, a similar mechanism seems unlikely since the gap (~4 eV [30]) is well above the transition energy.

In sum, the development of a new x-ray spectrometer allows investigation of electronic excitations with 25 meV resolution and acceptable count-rates at high momentum transfer using non-resonant inelastic x-ray scattering. We observe spectral changes in the band of *d-d* excitations in NiO at 1.7 eV energy transfer and |**Q**| = 6.52 Å$^{-1}$, with a complex line-shape at low temperature that broadens and becomes increasingly Gaussian as temperature is increased. Correction for the effect of thermal expansion, calculated to be -1.27% per pm of NiO lattice constant increase, leaves a remaining dependence that is consistent with the disappearance of magnetic order at $T_N$. Meanwhile the temperature dependence of the line width is in good agreement with a simple model for the Ni-O bond-length fluctuation, and yields a scale factor of 53(3) meV/pm fluctuation, in reasonable agreement with the 44.5 meV from calculation. This work demonstrates the potential of a new high-resolution spectrometer now coming on line at the RIKEN SPring-8 Center. It performs well with a single analyzer, and when expanded to its full complement of 9 analyzers, should be a formidable instrument for high-resolution NRIXS investigations of electronic excitations, with the T-Gradient design allowing a large area for complex sample environments.

E-mail: baron@spring8.or.jp, disikawa@spring8.or.jp, M.W.Haverkort@thphys.uni-heidelberg.de